%% file: vikram.tex
\documentclass[12pt,preprint]{aastex}

\shorttitle{Instabilities in Winds, SNe}
\shortauthors{Dwarkadas.}
\begin{document} 
\input{macros}
 
\title{Hydrodynamics of Supernova Evolution in the Winds of Massive Stars}

\author{ Vikram V. Dwarkadas}

\affil{Astronomy and Astrophysics, Univ of Chicago, 5640 S Ellis Ave
AAC 010c, Chicago IL 60637}
 
\email{vikram@oddjob.uchicago.edu}

\begin{abstract}
Core-Collapse supernovae arise from stars greater than 8
$\msun$. These stars lose a considerable amount of mass during their
lifetime, which accumulates around the star forming wind-blown
bubbles. Upon the death of the star in a spectacular explosion, the
resulting SN shock wave will interact with this modified medium. We
study the evolution of the shock wave, and investigate the properties
of this interaction. We concentrate on the evolution of the SN shock
wave in the medium around a 35 solar mass star. We discuss the
hydrodynamics of the resulting interaction, the formation and growth
of instabilities, and deviations from sphericity.
\end{abstract}

\keywords{supernova remnants; hydrodynamics; instabilities; stellar
winds; massive stars; wind-blown bubbles; shock waves}

\section{INTRODUCTION} 
Mass-loss from stars is a ubiquitous process. Massive stars ($> 8
\msun$) lose a considerable amount of mass before they explode. This
material collects around the star, forming a circumstellar (CS)
wind-blown bubble. At the end of its life, the star will explode in a
cataclysmic supernova (SN) explosion, and the resulting shock wave
will interact with this medium. The further evolution of the resulting
supernova remnant will depend on the properties of this medium.

In this paper we discuss the evolution of the surrounding medium
around massive stars, and the subsequent interaction of the SN shock
wave with this medium following the star's death. The rich and complex
dynamics of the various interactions leads to the formation and growth
of a variety of hydrodynamic instabilities, which we will focus on in
this paper. It may be possible to simulate some of these hydrodynamic
situations with available laboratory apparatus, and we hope that this
work will further stimulate laboratory experiments of realistic
astrophysical phenomena, particularly those involving radiative
shocks.

\section{SN-Circumstellar Interaction}

It has been realized over the years that the medium around a
core-collapse SN is continually being sculpted during the progenitor
star's lifetime, by the action of winds and outbursts. Chevalier \&
Liang (1989) discussed the interaction between a SN shock wave and the
surrounding wind-blown bubble formed by the pre-SN star. However
analytic arguments can only be extended so far, and numerical
simulations are required to study the subsequent non-linear
behavior. A series of papers in the early 90's (Tenorio-Tagle et
al.~1990, 1991; Rozyczka et al.~1993) explored some aspects of
this. Since then our observational knowledge of this phenomena has
multiplied exponentially, thanks to the availability of space based
data in the optical, X-ray and infrared bands, and the stream of data
pouring in from observations of SN 1987A. The latter has become the
poster-child for SN evolution in wind-blown bubbles, having shaped and
confirmed many of our views.

The basic details of SN interaction with wind-blown bubbles were
outlined in the papers listed above, and further elaborated on by
Dwarkadas (2005). The mass loss results in the formation of a
circumstellar (CS) wind blown cavity surrounding the star, bordered by
a thin, dense, cold shell. The typical structure of this wind-blown
bubble for constant wind properties (Weaver et al.~1977) is shown in
Figure 1, and consists of an outwards expanding shock wave (R$_o$),
and a wind termination shock (R$_t$) that expands inwards in a
Lagrangian sense, separated by a contact discontinuity (R$_{cd}$). In
general most of the volume between R$_t$ and R$_{cd}$ is occupied by a
low-density, high pressure shocked wind bubble, surrounded by the
extremely dense shell. Most of the mass is contained in the dense
shell. When the SN shock wave interacts with this bubble, it quickly
finds itself in a medium with density much lower than that of the
ISM. Consequently, the emission from the remnant, which arises mainly
from CS interaction (Chevalier \& Fransson 1994), will be considerably
reduced compared to evolution within the ISM.

\begin{figure}[htb]
\includegraphics[scale=0.8]{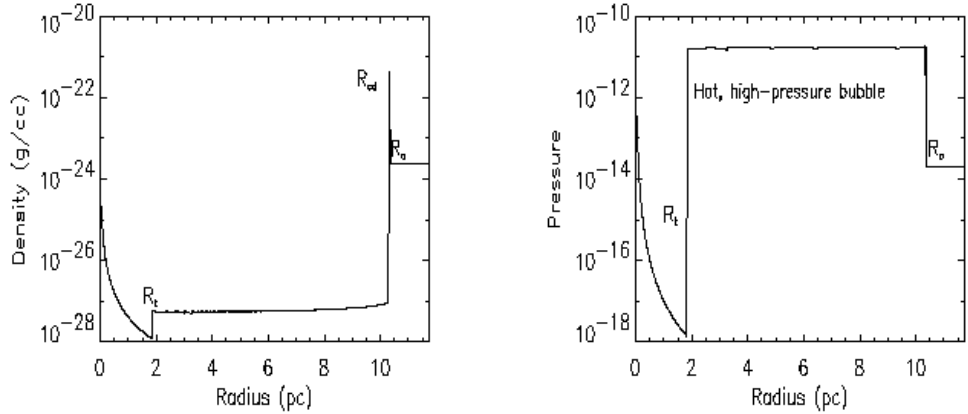}
\caption{Density and pressure profiles for a circumstellar wind-blown
bubble}
\label{fig:bub}
\end{figure}

It comes as no surprise then that the subsequent evolution depends
primarily on a single parameter $\Lambda$, the ratio of the mass of
the dense shell to that of the ejected material. For very small values
$\Lambda \ll 1$ the effect of the shell is negligible as expected. For
values of $\Lambda \la 1$ interaction with the shell results in
considerable deceleration of the SN shock wave. The X-ray luminosity
can increase by orders of magnitude upon shock-shell collision. A
transmitted shock wave enters the shell, while a reflected shock wave
moves back into the ejecta. If X-ray images were taken just after the
interaction, they would show the presence of a double-shelled
structure as the reflected shock begins to move inwards. In about 10
doubling times of the radius the SN begins to `forget' about the
existence of the shell. The remnant density profile changes to reflect
this, and consequently the X-ray emission from the remnant, which
depends on the density structure, will also change.  The reflected
shock will move to the center and presumably be reflected back, while
the transmitted shock will slowly exit the shell and eventually
separate from it.

As the ratio $\Lambda$ increases, more of the kinetic energy from the
remnant is converted to thermal energy of the shell. The transmitted
shock is considerably slowed down, and in extreme cases ($\Lambda \gg
1$) may even be trapped in the shell. The high pressure behind the
reflected shock will impart a large velocity to the shock, and
therefore thermalization of the ejecta is achieved in a much shorter
time as compared to thermalization by the SN reverse shock.  Upon
reaching the center the reflected shock bounces back, sending a weaker
shock wave that will collide with the shell. In time a series of shock
waves and rarefaction waves are seen traversing the ejecta. Each time
a shock wave collides with the dense shell a corresponding (but
successively weaker) rise in the X-ray emission from the remnant is
seen.

We have outlined the basics of SN interaction with CS wind-blown
shells. One-dimensional models are fully described in Dwarkadas
(2005). We wish to present herein results from multi-dimensional
models.

\section{CS Medium around a 35 $\msun$ Star}

The above description considered an idealized wind-blown bubble formed
by the interaction of a fast wind with the surrounding medium, where
the properties of both are constant in time. In reality, as a massive
star evolves, the wind properties change with time. In particular
after a star leaves the main sequence, its mass-loss properties change
considerably. This will give rise to a much more complicated bubble
structure than is shown in Figure 1.

In order to explore more realistically the medium surrounding a
core-collapse SN, we have taken stellar evolution calculations from
several groups, and investigated the evolution of the surrounding
medium as the star evolves. In this paper we discuss the evolution of
the medium around a 35 $\msun$ star, from an evolutionary model
provided to us by Norbert Langer. The star begins its life on the main
sequence as an O star, then expands to become a Red Supergiant (RSG),
and finally ends its life as a Wolf-Rayet (WR) star. The mass-loss
rate and wind velocity over the evolution are shown in Figure
2. During the main-sequence stage, the mass-loss rate is a few times
10$^{-7} \msun$/yr, and the wind velocity is about 3000-4000
km/s. Once the star swells to become a red supergiant, the wind
velocity reduces by more than 2 orders of magnitude, and the mass-loss
rate increases to almost 10$^{-4} \msun$/yr. The WR wind shows a
slight drop in mass-loss rates by a factor of a few from the RSG
phase, but a steep increase in wind velocity by two orders of
magnitude.  We use these values as input boundary values at each
timestep to our code, which then computes the structure of the nebula
over time. Unlike a previous computation (Garcia-Segura et al.~1996)
our computation is fully two-dimensional right from the start. We use
a grid consisting of 600 zones in both the radial and azimuthal
directions. The code used is the VH-1 numerical hydrodynamics code, a
multi-dimensional code that solves the equations of fluid dynamics on
a Lagrangian grid, and then remaps them onto an Eulerian
grid. Radiative cooling is included via a cooling function, but we
have not included the effects of ionization. A grid that expands
outwards with the outgoing shock wave is used, although no new zones
are added, i.e. the grid is not adaptive. Initially the wind occupies
about 20 zones on the grid, depending on the grid resolution. The
mass-loss rate and velocity of the wind are used to compute the
density and velocity of the inflow, which are used as the input
boundary conditions at each timestep. The initial setup is uniform and
no perturbations are applied to the system. Perturbations that arise
are due to effects such as non-spherical shocks on the spherical grid.

\begin{figure}[htb]
\includegraphics[scale=1.]{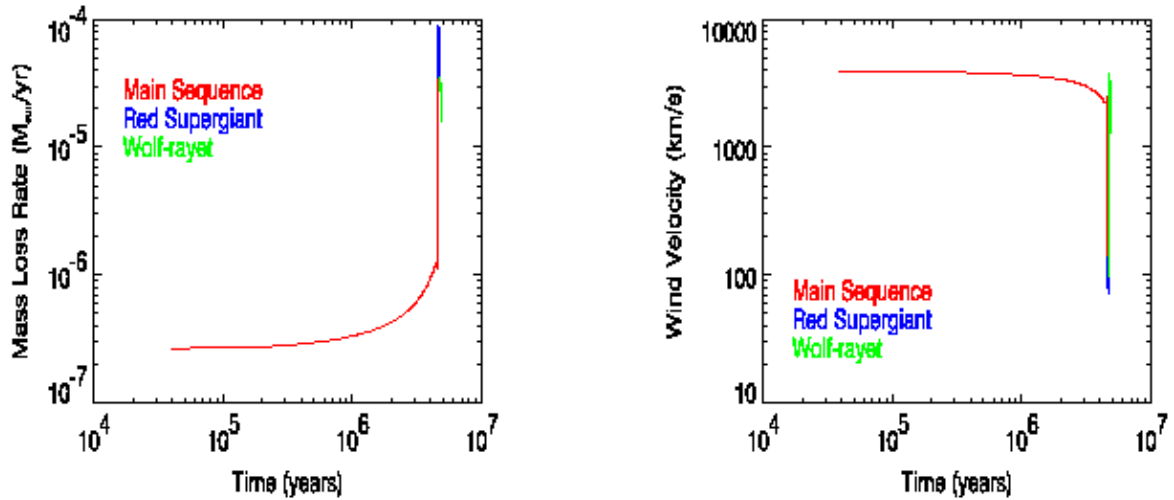}
\caption{Evolution of the mass-loss rate (left) and wind velocity
(right) for the 35 $\msun$ star during its lifetime}
\label{fig:params}
\end{figure}

The evolution of the medium is shown in Figure 3. In the main sequence
stage (3a, 3b), although the wind properties are changing
continuously, the nebular structure is not very different from that
expected from the idealized, two-wind case of a fast wind interacting
with a slower wind, both of which have constant wind properties. A
thin shell of swept-up material is formed, and the volume of the
nebula is mostly occupied by a hot, low density bubble. The shell is
on the whole mostly stable, although it shows some wrinkles. These
arise mainly from shearing, due to flow of gas along the contact
discontinuity. However these instabilities are not highly pronounced,
and do not appear to grow to any significant extent. In
lower-resolution simulations presented earlier (Dwarkadas 2001, 2004)
we had suggested that the shell is unstable to some type of thin-shell
instability (Vishniac 1983). We have seen the same instability in
simulations of the medium around a 40 $\msun$ star. The higher
resolution calculations presented here do not show a strong presence
of such an instability however. This is a topic still under
investigation.

\begin{figure}[htb]
\includegraphics[scale=0.85]{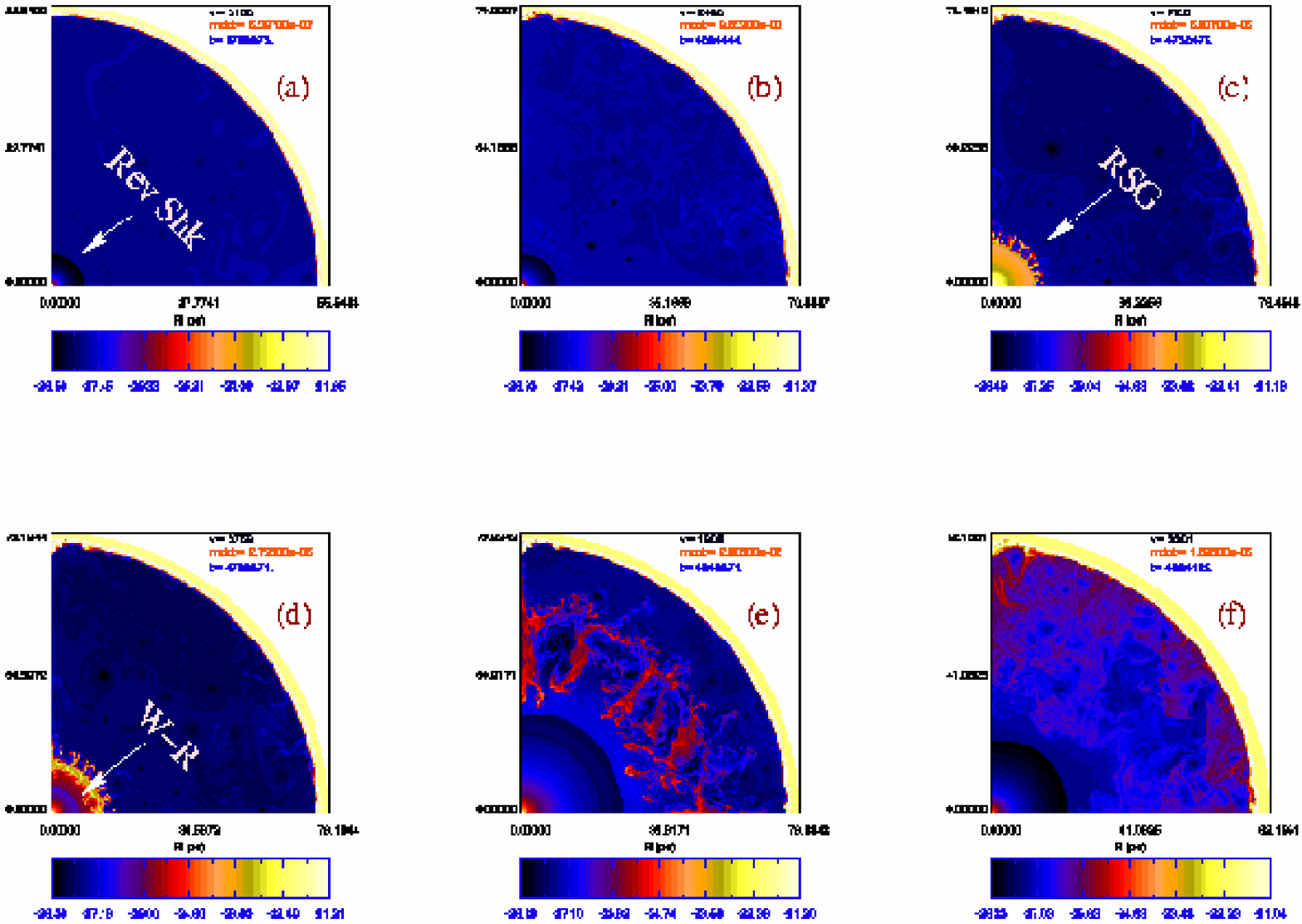}
\caption{Density evolution of the medium a round a 35 $\msun$ star
with time. The wind properties at each stage are given in the top
right hand corner of each panel. The velocity is in km/s, the mass
loss rate in $\msun$/yr, and the time in years. The color bar shows
the logarithm of the gas density in units of g cc$^{-1}$}
\label{fig:wrbub}
\end{figure}

The interior of the nebula shows significant fluctuations in density
and pressure, and vortices are visible in the velocity flow. Since the
mass-loss rate and wind velocity are changing at every timestep, the
position of the reverse shock is not fixed on the grid, but moves
slightly every timestep with respect to the outer shock. The changing
position of the reverse shock from one timestep to another results in
the deposition of vorticity into the shocked wind, which is then
carried out with the shocked flow. This results in an interior that is
quite inhomogeneous, with significant density fluctuations.

When the star leaves the main sequence and becomes a RSG star, its
radius increases considerably, the wind velocity ($V_{wind}$) drops by
two orders of magnitude, while the mass-loss rate ($\dot{M}$)
increases appropriately. Thus the wind density, proportional to
$\dot{M}/V_{wind}$, goes up by several orders of magnitude. A new
pressure equilibrium is established, and a shock front is formed in
between the RSG wind and main sequence bubble as the RSG wind is
decelerated by the bubble pressure. The RSG wind piles up against this
shock, forming a thin dense shell of RSG material. No hot, low-density
cavity is present. The shell decelerates as it expands outwards,
satisfying the classic case of Rayleigh-Taylor (R-T) instability, and
{\it Rayleigh-Taylor fingers are seen expanding outwards from the
high-density shell into the low-density ambient medium} (3c).  Some of
the filaments show the presence of sub-filaments growing from the main
one, and the expanded heads of many of the filaments are a sign of
Kelvin-Helmholtz instabilities resulting from the shear flow in
between the filaments and the surrounding medium. Unfortunately, since
we need to resolve the entire bubble, the resolution is not large
enough to study the growth of the fingers in detail.

The star leaves the RSG phase and loses its outer hydrogen envelope,
becoming a WR star in the process. The compact star now gives off a
very fast wind, not unlike in the O star stage, but with a mass-loss
rate that is much higher than in the main sequence, and just a few
times lower than the RSG stage. The supersonic WR wind creates a
wind-blown bubble in the RSG wind. The dense W-R shell is accelerated
by the high pressure, low density interior as it expands outwards,
leading to the triggering of the Rayleigh-Taylor instability (3d). In
this case R-T fingers are seen expanding {\it inwards from the dense
shell into the low-density cavity}. The large momentum of the WR shell
causes the RSG shell to fragment, and carries the material outwards
(3e), speeding up as it enters the low-density bubble. Due to the fact
that the WR wind is carrying fragments of the RSG material, and that
it travels through a medium with considerable fluctuations in density
and pressure, its expansion is not completely spherical. The collision
of this slightly aspherical wind with the main sequence shell gives
rise to a reflected shock that moves back into the bubble. The
asphericity is accentuated in the reflected shock, which moves
inwards, before finally coming to rest in a wind-termination shock
where the ram pressure of the freely expanding wind is equal to the
thermal pressure within the bubble. The wind-termination shock when it
forms is consequently also not spherical but slightly elongated along
the equator (3f). As we shall show later this has important
consequences for the expansion of the SN shock wave.

\section{SN-CSM interaction in the case of the 35 $\msun$ star}

At the end of the WR stage, the stellar mass remaining is 9.1
$\msun$. We assume that the star then explodes in a SN explosion. A
remnant of 1.4 $\msun$ is left behind, and the remaining mass is
ejected in the explosion. We use the prescription of Chevalier \&
Fransson (1994) to describe the ejecta structure as a power-law with
density, with power-law index of 7. We compute the evolution of the SN
described by this density profile expanding into the unshocked wind,
and then map it onto the grid containing the bubble simulation. This
calculation was also carried out using 600 X 600 zones.

The evolution of the SN shock wave is shown in Figure 4. It starts out
as expected, with the formation of a forward and reverse shock
structure (Fig 4a). The interaction of the spherical forward shock
with the aspherical wind termination shock, susceptible to the
Richtmeyer-Meshkov instability, reveals quite interesting
dynamics. Since the SN shock is spherical while the wind termination
shock is slightly more elongated towards the equator, the interaction
first takes place close to the symmetry axis. A transmitted shock
moves out into the shocked bubble, while a reflected shock moves back.
Different parts of the SN shock collide with the wind-termination
shock at different times, leading to transmitted shocks with a small
but non-negligible velocity spread (Fig 4b). The composite transmitted
shock then expands in the inhomogeneous medium, interacting with
several large density fluctuations on the way. The net result is a
very corrugated shock wave that expands outwards towards the main
sequence shell (Fig 4c). The wrinkles are similarly prevalent in the
reverse shock also. The wrinkled shock collides in a piecemeal fashion
with the main sequence shell, with some parts of the shock colliding
before others (Fig 4d, 4e). Each collision with the shell will give
rise to an increase in the optical and X-ray emission at that
point. Therefore some parts of the shell will brighten before
others. It is interesting to note that a similar phenomenon has been
observed in SN 1987A, where bright spots appear successively around
different parts of the equatorial ring (Sugerman et al.~2002)

As each portion of the forward shock wave collides with the shell, a
reflected and transmitted shock pair is formed. The shell is dense
enough that the transmitted shock does not emerge from the shell for a
long period. The reflected shock meanwhile travels back towards the
origin. However, as seen in Figure 4, the velocity of each piece
differs considerably from the next, both in magnitude as well as
direction. The shape of the reflected shock therefore deviates
significantly from spherical, and some parts of the reflected shock
reach the symmetry axis before the rest has traveled far into the
interior (Figure 4f). This gives the remnant a very asymmetric shape,
and results in some portions of the ejecta being much hotter than
others.

\section{Conclusions}

The surroundings of massive stars are shaped by the mass-loss from the
progenitor star. This can lead to a complicated density structure for
the surrounding medium, the formation and growth of various
hydrodynamical instabilities, deposition of vorticity and onset of
turbulence. When the star explodes as a SN remnant, the SN shock wave
will interact with this ambient medium. The inhomogeneous structure of
the ambient medium can cause distortions in the SN shock wave as it
expands outwards, which are magnified by the turbulence in the
wind-blown structure. In this paper we have shown that the end result
can be a wrinkled shock wave whose impact with the surrounding shell
occurs in a piecemeal fashion. As each part of the shock wave hits the
shell, it will brighten up in the optical and X-ray regime, a
phenomenon that is observable in SN 1987A. We do caution that this
comparison is illustrative only. Our numerical models are not meant to
simulate SN 1987A, whose progenitor star was a much lower mass B3Ia
star.

\begin{figure}[htb]
\includegraphics[scale=0.85]{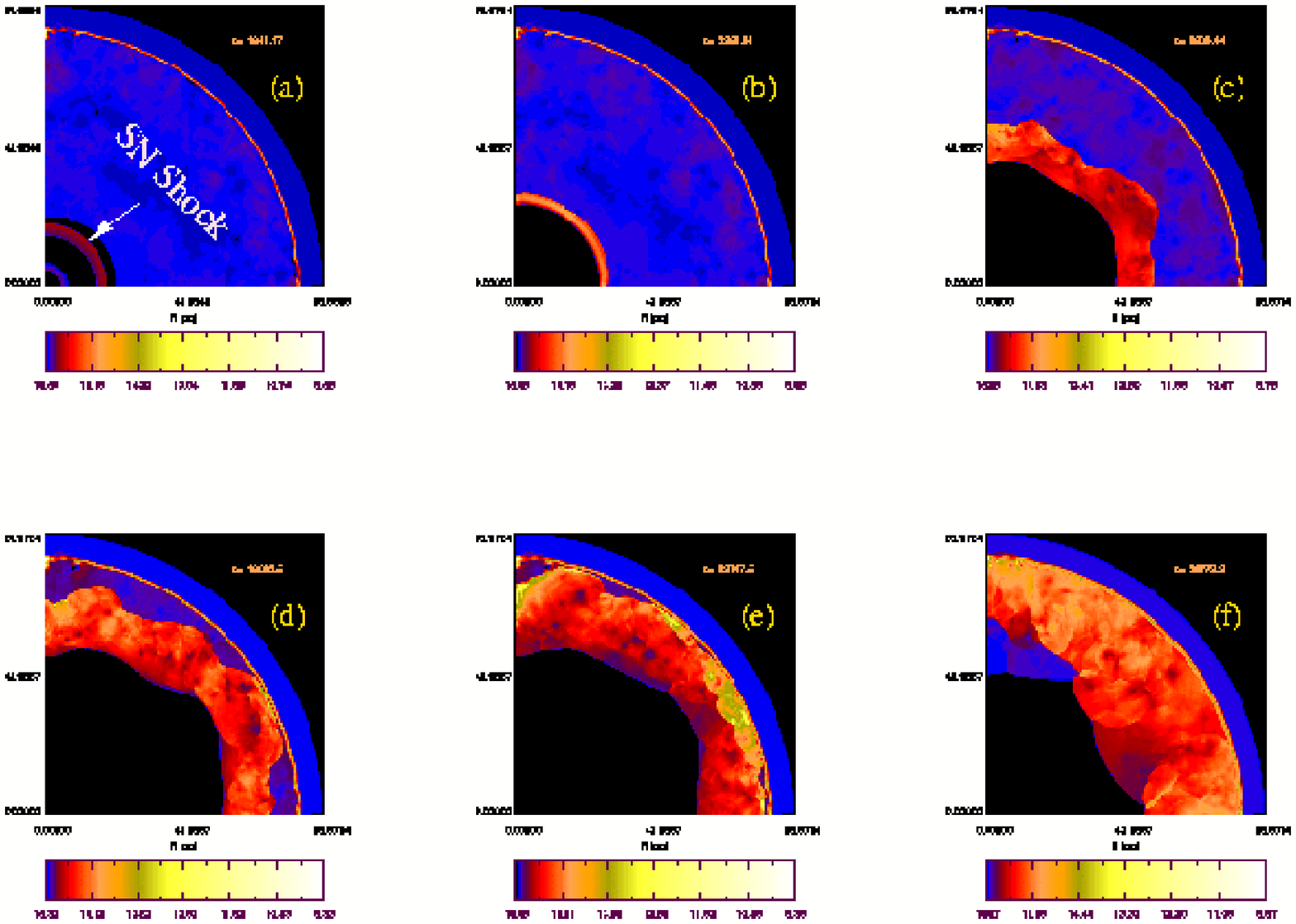}
\caption{Pressure evolution of the SN shock wave within the WR bubble}
\label{fig:snbub}
\end{figure}

Our simulations show that the complicated structure of the medium may
result in deviations from spherical symmetry for the SN shock
wave. Even though the expansion starts out as spherical, the final
shape of the remnant may deviate considerably from sphericity. Most of
the emission from the remnant arises from the high pressure region in
between the forward and reverse shocks. The distorted shape of this
emitting region is clearly visible in Figure 4, and this will be
reflected in observations of the remnant.

Herein we have summarized the features of multi-dimensional models of
SN evolution in the environments shaped by massive stars. Further
details are available from Dwarkadas (2006).

Vikram Dwarkadas is supported by award \# AST-0319261 from the
National Science Foundation, and by NASA through grant \# HST-AR-10649
from STScI. We thank the anonymous referee for suggestions that helped
to improve this paper. We acknowledge useful discussions with Roger
Chevalier which were particularly helpful in identifying the various
instabilities that were observed. We are grateful for comments from
John Blondin and Thierry Foglizzo.  This research was supported in
part by the National Science Foundation under Grant No. PHY99-07949 to
the KITP.

\end{document}

%% file: macros.tex
\newcommand{\vper}{\mbox{${v_{\perp}}$}}
\newcommand{\vpar}{\mbox{${v_{\parallel}}$}}
\newcommand{\uper}{\mbox{${u_{\perp}}$}}
\newcommand{\vperout}{\mbox{${{v_{\perp}}_{o}}$}}
\newcommand{\uperout}{\mbox{${{u_{\perp}}_{o}}$}}
\newcommand{\vperin}{\mbox{${{v_{\perp}}_{i}}$}}
\newcommand{\uperin}{\mbox{${{u_{\perp}}_{i}}$}}
\newcommand{\upar}{\mbox{${u_{\parallel}}$}}
\newcommand{\uparout}{\mbox{${{u_{\parallel}}_{o}}$}}
\newcommand{\vparout}{\mbox{${{v_{\parallel}}_{o}}$}}
\newcommand{\uparin}{\mbox{${{u_{\parallel}}_{i}}$}}
\newcommand{\vparin}{\mbox{${{v_{\parallel}}_{i}}$}}
\newcommand{\dout}{\mbox{${\rho}_{o}$}}
\newcommand{\din}{\mbox{${\rho}_{i}$}}
\newcommand{\da}{\mbox{${\rho}_{1}$}}
\newcommand{\mfast}{\mbox{$\dot{M}_{f}$}}
\newcommand{\mslow}{\mbox{$\dot{M}_{a}$}}
\newcommand{\beqn}{\begin{eqnarray}}
\newcommand{\eeqn}{\end{eqnarray}}
\newcommand{\be}{\begin{equation}}
\newcommand{\ee}{\end{equation}}
\newcommand{\noi}{\noindent}
\newcommand{\ftheta}{\mbox{$f(\theta)$}}
\newcommand{\gtheta}{\mbox{$g(\theta)$}}
\newcommand{\ltheta}{\mbox{$L(\theta)$}}
\newcommand{\stheta}{\mbox{$S(\theta)$}}
\newcommand{\utheta}{\mbox{$U(\theta)$}}
\newcommand{\xitheta}{\mbox{$\xi(\theta)$}}
\newcommand{\vs}{\mbox{${v_{s}}$}}
\newcommand{\ro}{\mbox{${R_{0}}$}}
\newcommand{\pa}{\mbox{${P_{1}}$}}
\newcommand{\va}{\mbox{${v_{a}}$}}
\newcommand{\vo}{\mbox{${v_{o}}$}}
\newcommand{\vp}{\mbox{${v_{p}}$}}
\newcommand{\vw}{\mbox{${v_{w}}$}}
\newcommand{\vf}{\mbox{${v_{f}}$}}
\newcommand{\lprime}{\mbox{${L^{\prime}}$}}
\newcommand{\uprime}{\mbox{${U^{\prime}}$}}
\newcommand{\sprime}{\mbox{${S^{\prime}}$}}
\newcommand{\xiprime}{\mbox{${{\xi}^{\prime}}$}}
\newcommand{\mdot}{\mbox{$\dot{M}$}}
\newcommand{\msun}{\mbox{$M_{\odot}$}}
\newcommand{\yr}{\mbox{${\rm yr}^{-1}$}}
\newcommand{\kms}{\mbox{${\rm km} \;{\rm s}^{-1}$}}
\newcommand{\lambdav}{\mbox{${\lambda}_{v}$}}
\newcommand{\lequ}{\mbox{${L_{eq}}$}}
\newcommand{\eqpratio}{\mbox{${R_{eq}/R_{p}}$}}
\newcommand{\ra}{\mbox{${r_{o}}$}}
\newcommand{\bfig}{\begin{figure}[h]}
\newcommand{\efig}{\end{figure}}
\newcommand{\tone}{\mbox{${t_{1}}$}}
\newcommand{\done}{\mbox{${{\rho}_{1}}$}}
\newcommand{\dsn}{\mbox{${\rho}_{SN}$}}
\newcommand{\dzero}{\mbox{${\rho}_{0}$}}
\newcommand{\ve}{\mbox{${v}_{e}$}}
\newcommand{\vej}{\mbox{${v}_{ej}$}}
\newcommand{\Mch}{\mbox{${M}_{ch}$}}
\newcommand{\mej}{\mbox{${M}_{e}$}}
\newcommand{\Mst}{\mbox{${M}_{ST}$}}
\newcommand{\dam}{\mbox{${\rho}_{am}$}}
\newcommand{\Rst}{\mbox{${R}_{ST}$}}
\newcommand{\Vst}{\mbox{${V}_{ST}$}}
\newcommand{\Tst}{\mbox{${T}_{ST}$}}
\newcommand{\no}{\mbox{${n}_{0}$}}
\newcommand{\Efif}{\mbox{${E}_{51}$}}
\newcommand{\rsh}{\mbox{${R}_{sh}$}}
\newcommand{\msh}{\mbox{${M}_{sh}$}}
\newcommand{\vsh}{\mbox{${V}_{sh}$}}
\newcommand{\vrev}{\mbox{${v}_{rev}$}}
\newcommand{\rpr}{\mbox{${R}^{\prime}$}}
\newcommand{\mpr}{\mbox{${M}^{\prime}$}}
\newcommand{\vpr}{\mbox{${V}^{\prime}$}}
\newcommand{\tpr}{\mbox{${t}^{\prime}$}}
\newcommand{\cone}{\mbox{${c}_{1}$}}
\newcommand{\ctwo}{\mbox{${c}_{2}$}}
\newcommand{\cthree}{\mbox{${c}_{3}$}}
\newcommand{\cfour}{\mbox{${c}_{4}$}}
\newcommand{\Te}{\mbox{${T}_{e}$}}
\newcommand{\Ti}{\mbox{${T}_{i}$}}
\newcommand{\Ha}{\mbox{${H}_{\alpha}$}}
\newcommand{\Rprime}{\mbox{${R}^{\prime}$}}
\newcommand{\Vprime}{\mbox{${V}^{\prime}$}}
\newcommand{\Tprime}{\mbox{${T}^{\prime}$}}
\newcommand{\Mprime}{\mbox{${M}^{\prime}$}}
\newcommand{\rprime}{\mbox{${r}^{\prime}$}}
\newcommand{\rfprime}{\mbox{${r}_f^{\prime}$}}
\newcommand{\vprime}{\mbox{${v}^{\prime}$}}
\newcommand{\tprime}{\mbox{${t}^{\prime}$}}
\newcommand{\mprime}{\mbox{${m}^{\prime}$}}
\newcommand{\Me}{\mbox{${M}_{e}$}}
\newcommand{\nh}{\mbox{${n}_{H}$}}
\newcommand{\rr}{\mbox{${R}_{2}$}}
\newcommand{\rf}{\mbox{${R}_{1}$}}
\newcommand{\vtwo}{\mbox{${V}_{2}$}}
\newcommand{\vout}{\mbox{${V}_{1}$}}
\newcommand{\dshell}{\mbox{${{\rho}_{sh}}$}}
\newcommand{\dwind}{\mbox{${{\rho}_{w}}$}}
\newcommand{\dslow}{\mbox{${{\rho}_{s}}$}}
\newcommand{\dfast}{\mbox{${{\rho}_{f}}$}}
\newcommand{\vfast}{\mbox{${v}_{f}$}}
\newcommand{\vslow}{\mbox{${v}_{s}$}}
\newcommand{\cc}{\mbox{${\rm cm}^{-3}$}}